\documentclass[prb,twocolumn]{revtex4-1}
\usepackage{times}
\usepackage{bbm}
\usepackage[dvips]{graphicx}
\usepackage{latexsym,amsmath,amssymb,bm,euscript,eufrak}
\usepackage[dvips]{color}
\usepackage{multirow}
\usepackage{calligra}
\usepackage{color}
\usepackage[colorlinks=true,linkcolor=blue,citecolor=blue]{hyperref}
\usepackage{hyperref}

\def\uu{B}
\def\atil{\tilde{a}_\mu(r)}
\def\ab{a_\mu(r)}
\def\cQ{{ Q}}

\def\hext{h^{\rm ext}}

\begin{document}

\title{Model of Fractionalization of Faraday Lines in Compact Electrodynamics}
\date{\today}

\author{Scott D. Geraedts}
\author{Olexei I. Motrunich}
\affiliation{Department of Physics, California Institute of Technology, Pasadena, California 91125, USA}

\begin{abstract}
Motivated by ideas of fractionalization and intrinsic topological order in bosonic models with short-range interactions, we consider similar phenomena in formal lattice gauge theory models. Specifically, we show that a compact quantum electrodynamics (CQED) can have, besides the familiar Coulomb and confined phases, additional unusual confined phases where excitations are quantum lines carrying fractions of the elementary unit of electric field strength. We construct a model that has $N$-tupled monopole condensation and realizes $1/N$ fractionalization of the quantum Faraday lines.  This phase has another excitation which is a $Z_N$ quantum surface in spatial dimensions five and higher, but can be viewed as a quantum line or a quantum particle in four or three spatial dimensions respectively.  These excitation have statistical interactions with the fractionalized Faraday lines; for example, in three spatial dimensions, the particle excitation picks up a Berry phase of $e^{i2\pi/N}$ when going around the fractionalized Faraday line excitation. We demonstrate the existence of this phase by Monte Carlo simulations in (3+1) space-time dimensions.
\end{abstract}
\maketitle

\section{Introduction}

The classification of the topological phases of gauge theories is a longstanding problem.\cite{KapustinThorngren, GukovKapustin} The study of such phases can be made easier if they can be realized in a precisely defined lattice model.  In this work we provide a lattice model which realizes a topological phase of compact electrodynamics which confines electrical charges but shows fractionalized Faraday line excitations. 

Ordinary compact quantum electrodynamics (CQED) has two phases: a confined phase and a Coulomb phase. We can think about these phases in terms of the behavior of the monopoles in the system. In the confined phase monopoles are condensed (the system contains many monopoles), while in the Coulomb phase monopoles are gapped (the system contains few monopoles). In our model we consider the case where individual monopoles are gapped, but bound states of $N$ monopoles are condensed, which we argue leads to fractionalized Faraday lines.

Such an approach is inspired by the search for topological phases in condensed matter physics, where it is known that the condensation of multiple topological defects can lead to fractionalized phases with intrinsic topological order.\cite{BalentsFisherNayak99, SenthilFisher00}  

For example, in (2+1) dimensions a system of bosons can be reformulated in terms of vortices, which are quantum particles. The (dual) field theory for the vortices has the structure of a Higgs theory with vortex fields minimally coupled to a dynamical gauge field, whose flux is the coarse-grained boson density. When the vortex fields condense we have a Higgs phase, which in the boson language is simply a Mott insulator. The Higgs phase contains Abrikosov-Nielsen-Olesen (ANO) vortices, which are topological defects of the original vortex fields (as opposed to the original vortices, which are topological defects of the boson fields).  The ANO vortices in the dual theory are gapped excitations in this phase and can be identified with gapped charge excitations in the Mott insulator of the bosons.  Since the ANO vortices in the Higgs phase carry quantized flux, the boson charge is quantized in the Mott insulator.

We can then consider condensing pairs of vortices while leaving single vortices gapped.\cite{SenthilFisher00}  This is still a Mott insulator of bosons, but an unusual one.  ANO vortices of the `paired-vortex' field have flux quantized to half-integers, and so they correspond to gapped charge excitations which carry half a unit of the fundamental boson charge.  Original single vortices are also gapped excitations acquiring $Z_2$ character, since the paired-vortex condensate can absorb any even number of vortices. Furthermore, a fractionalized charge picks up a phase of $\pi$ when taken around a single vortex.  Therefore the phase where pairs of vortices are condensed is a fractionalized Mott insulator whose excitations are particles carrying half a unit of charge (which we will call 'chargons') and $Z_2$ fluxes carrying no charge and exhibiting mutual statistics with the chargons.  A convenient mathematical description for this phase has bosonic chargons coupled to a $Z_2$ gauge field in the deconfined phase, and one of the signatures of the topological nature of the phase is the ground state degeneracy of $2^d$ when the system is placed on a torus in $d$ dimensions.

Turning now to CQED in (3+1)D, it can be formulated in terms of monopoles, which are also quantum particles that can be described by a Higgs theory.\cite{Polyakov} ANO vortices of the monopole fields in this case are quantum lines. The Higgs phase in the dual theory where the monopoles condense corresponds to the confining phase of the CQED, and the ANO vortices in the Higgs theory are the Faraday lines in the CQED, which are the gapped excitations of the confining phase. The quantization of the flux carried by the ANO vortices now leads to the quantization of electric field strength for the Faraday lines.

When bound states of $N$ monopoles condense, the flux of ANO vortices in the `$N$-tupled-monopole' field is quantized in units of $1/N$, and therefore they correspond to fractionalized Faraday (electric field) lines. The model we describe here is inspired by this argument,\cite{GukovKapustin} though we will explicitly demonstrate that our model produces fractionalized phases in all space-time dimensions greater than or equal to four. The model works by energetically binding together multiple monopoles, and we also study it numerically in (3+1) dimensions using Monte Carlo.  We will also derive microscopically an effective description of this phase in terms of fractionalized Faraday lines coupled to a rank-2 $Z_N$ field (``Gerbe field'') in the rank-2 deconfined phase, and we will show the ground state degeneracy of $N^{d(d-1)/2}$ on a $d$-dimensional torus.

\section{Model and Monte Carlo Study}
Our model is described by the following action:
\begin{widetext}
\begin{eqnarray}
Z = \int_0^{2\pi} Da_\mu(r) \sum_{\uu_{\mu\nu}(r) = -\infty}^\infty
\exp\left(-\frac{\kappa}{2} \sum_{r,~ \mu < \nu} \left[(\nabla_\mu a_\nu -\nabla_\nu a_\mu)(r) - 2\pi \uu_{\mu\nu}(r) \right]^2
+ \lambda \sum_{r,~ \sigma < \mu < \nu} \cos\left[\frac{2\pi Q_{\sigma\mu\nu}(r)}{N} \right] \right) ~.
\label{action1}
\end{eqnarray}
\end{widetext}
Here $a_\mu(r)$ are $2\pi$-periodic (compact) gauge fields, which live on the links of a (hyper)cubic lattice whose positions are labeled by $r$ and space-time directions by $\mu,\nu$, etc.; $\uu_{\mu\nu}(r)$ are integer valued variables which live on the plackets of the same lattice. Lattice derivatives are represented by $\nabla_\mu$. In the absence of the $\lambda$-term, the above is the familiar Villain form of the compact electrodynamics;\cite{Polyakov} in particular, summation over $\uu_{\mu\nu}(r)$ on each placket gives a $2\pi$-periodic ``Villain cosine'' potential on the gauge field fluxes:
\begin{equation}
e^{-V_{\rm Villain}[\Phi; \kappa]} \equiv
\sum_{\uu = -\infty}^\infty e^{-\frac{\kappa}{2} (\Phi - 2\pi \uu)^2} ~
=\sum_{F=-\infty}^{\infty} e^{-\frac{F^2}{2\kappa}+iF\Phi} ~.
\label{Villain}
\end{equation}
The last equality comes from Poisson resummation and will be used later (and we dropped unimportant constant factors). Keeping $\uu_{\mu\nu}(r)$ dynamical allows us to define ``monopolicities''\cite{Polyakov} associated with cubes defined on directions $\sigma < \mu < \nu$:
\begin{equation}
Q_{\sigma\mu\nu}(r) = (\nabla_\sigma \uu_{\mu\nu} + \nabla_\mu \uu_{\nu\sigma} + \nabla_\nu \uu_{\sigma\mu})(r) ~.
\label{Qsmn}
\end{equation}

As an example, in (3+1)D we can alternatively view these variables as objects residing on the dual lattice links, which we can then interpret as monopole currents,
\begin{equation}
J^{(m)}_\rho = \frac{1}{6} \sum_{\sigma\mu\nu} \epsilon_{\rho\sigma\mu\nu} Q_{\sigma\mu\nu}
= \frac{1}{2} \sum_{\sigma\mu\nu} \epsilon_{\rho\sigma\mu\nu} \nabla_\sigma \uu_{\mu\nu} ~. 
\label{Jdef}
\end{equation}
This provides a useful connection to the discussion in terms of the dual Higgs model in the Introduction, which utilized intuition about such Higgs models.  However, this is not required for the treatment in the next section which works in general dimensionality.

\begin{figure}
\includegraphics[angle=-90,width=\linewidth]{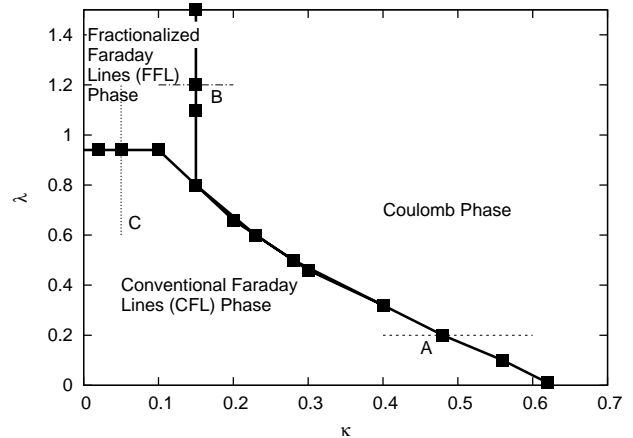}
\caption{Phase diagram for the generalized CQED model in Eq.~(\ref{action1}), which realizes a novel phase with fractionalized Faraday lines at large $\lambda$ and small $\kappa$.  The other phases are a conventional confined phase with gapped, conventional Faraday lines and Coulomb phase with a gapless photon and condensed Faraday lines; these phases are familiar from the ordinary CQED in (3+1) dimensions. Points indicate where the locations of the phase transitions were determined from peaks in the specific heat. Dashed lines indicate where more detailed data was taken, which is presented below.}
\label{phase}
\end{figure}

We study this model in Monte Carlo for $N=2$\cite{footnote1} in (3+1)D and show the numerical phase diagram in Fig.~\ref{phase}.  In the model, $\kappa$ represents the ``bare stiffness'' of the gauge field, while $\lambda$ represents the strength of the potential which penalizes single monopole excitations compared to pairs of monopoles. When $\lambda$ is small we recover the phase diagram of the ordinary CQED. 
At small $\kappa$ we have a confined phase, in which the monopoles are condensed. We will call this phase the ``Conventional Faraday Lines'' (CFL) phase, because its gapped excitations are conventional Faraday lines. At large $\kappa$ we have the Coulomb phase where monopoles are gapped and Faraday lines condense. This phase also has a gapless photon. As $\lambda$ is increased we find a new phase at large $\lambda$ and small $\kappa$, which we claim is a novel confined phase whose gapped excitations are fractionalized Faraday lines, and so we call it the ``Fractionalized Faraday Lines''(FFL) phase. 

When studying the above action in Monte Carlo, we measure specific heat, defined as 
\begin{equation}
C\equiv \frac{\langle S^2 \rangle -\langle S \rangle^2}{{\rm Vol}},
\end{equation}
where ${\rm Vol}=L^4$ is the volume of the system with linear dimension $L$. Peaks in the specific heat can be used to detect phase transitions even without knowing the nature of the phases. 

We also measure ``photon stiffness,''
\begin{eqnarray}
\rho_{\mu\nu}(k)&\equiv& \kappa - \kappa^2 \langle \left|\omega_{\mu\nu}(k)\right|^2 \rangle ~,
\label{rho}
\end{eqnarray}
where
\begin{eqnarray}
\omega_{\mu\nu}(r)&\equiv& (\nabla_\mu a_\nu -\nabla_\nu a_\mu)(r) -2\pi\uu_{\mu\nu}(r).
\end{eqnarray}
To obtain the above expression for the stiffness we couple the CQED system to an external, probing rank-2 field $\hext_{\mu\nu}(r)$ by making the following substitution in Eq.~(\ref{action1}):\cite{Polyakov}
\begin{equation}
\omega_{\mu\nu}(r) \rightarrow \omega_{\mu\nu}(r) - \hext_{\mu\nu}(r) ~.
\label{sub}
\end{equation} 
Just as superfluid stiffness in a system of bosons can be represented as a second derivative of a free energy (i.e., $-\log Z$) with respect to a probing field coupled to bosons, the photon stiffness Eq.~(\ref{rho}) can be derived by taking the second derivative of the free energy for our model Eq.~(\ref{action1}) with respect to $\hext_{\mu\nu}$. We measure the photon stiffness at the smallest non-zero wave-vector $k=k_{\rm min}$, which points in either the $\mu$ or $\nu$ directions and has magnitude $2\pi/L$. Note that strictly speaking when we use the substitution in Eq.~(\ref{sub}), we should also make the substitution 
\begin{equation}
Q_{\sigma\mu\nu}\rightarrow Q_{\sigma\mu\nu} +
\frac{\nabla_\sigma \hext_{\mu\nu} + \nabla_\mu \hext_{\nu\sigma} + \nabla_\nu \hext_{\sigma\mu}}{2\pi}
\end{equation}
in the $\lambda$-term of Eq.~(\ref{action1}), but the contribution to the stiffness from this term already contains derivatives and is proportional to $k^2$. When $k=k_{\rm min}$, this contribution vanishes at large $L$, and so we neglect it from now on. The photon stiffness should be non-zero only in the Coulomb phase, because it is the only phase with a gapless photon. Measuring vanishing photon stiffness then tells us that a phase is confined.

\begin{figure}[t]
\includegraphics[angle=-90,width=\linewidth]{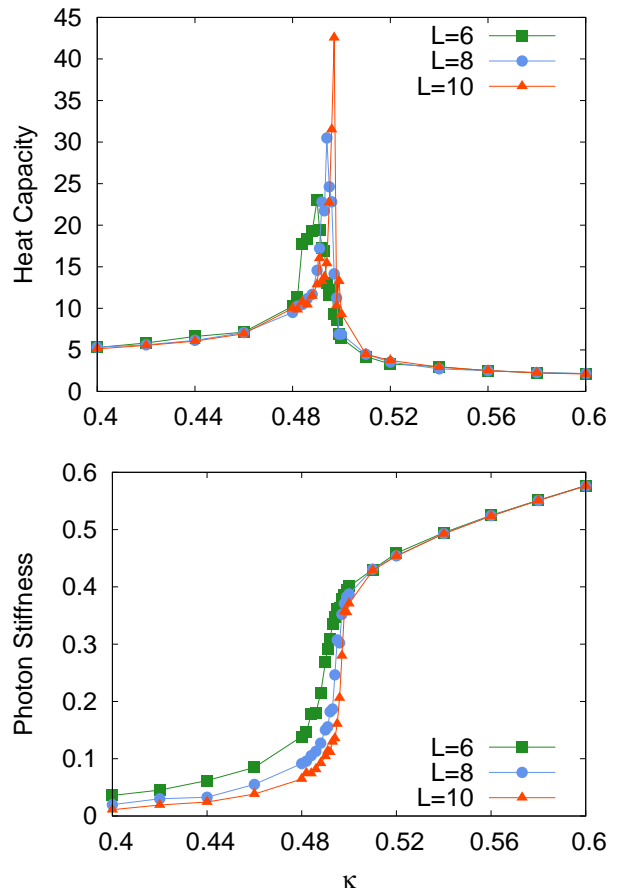}
\caption{Specific heat (top panel) and photon stiffness (bottom panel) along the line marked A in Fig.~\ref{phase}. The rapidly growing peak in the specific heat shows that we have a first-order transition, as expected. The photon stiffness shows that the phase at large $\kappa$ is the Coulomb phase, while the phase at small $\kappa$ is confined.}
\label{cutA}
\end{figure}

\begin{figure}[t]
\includegraphics[angle=-90,width=\linewidth]{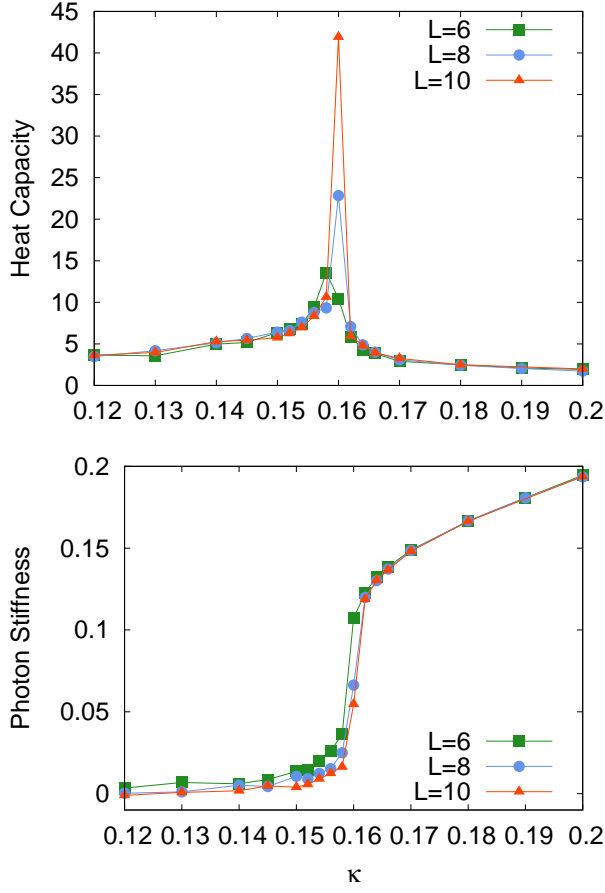}
\caption{Same as Fig.~\ref{cutA}, but for the line marked B in Fig.~\ref{phase}. The phase transition along this line, from the Coulomb to FFL phases, involves condensation of pairs of monopoles and is also described by a Higgs-like theory.  Therefore it has similar first-order behavior as the transition in Fig.~\ref{cutA} which corresponds to condensation of single monopoles.}
\label{cutB}
\end{figure}

\begin{figure}[t]
\includegraphics[angle=-90,width=\linewidth]{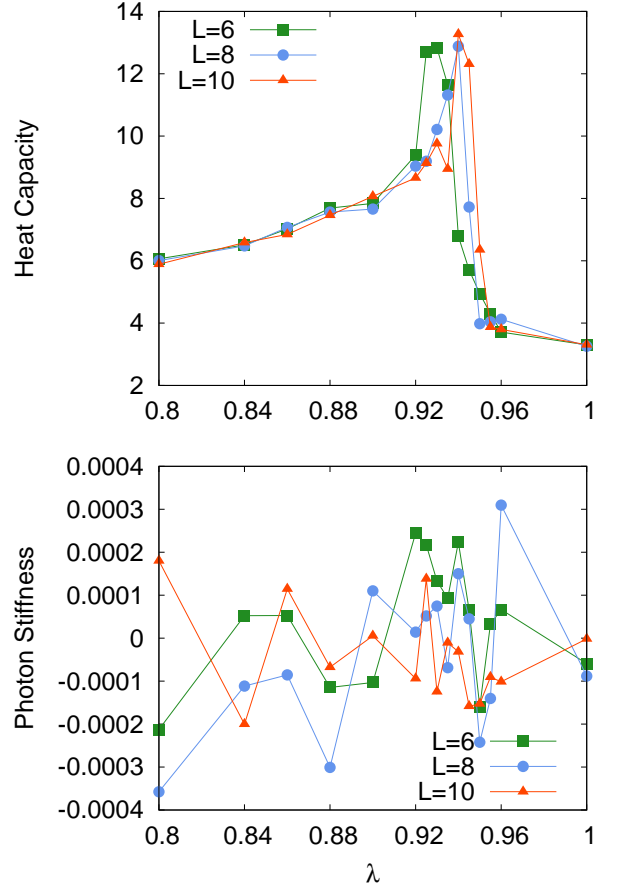}
\caption{Same as Fig.~\ref{cutA}, but for the line marked C in Fig.~\ref{phase}. The peak in the specific heat tells us that we have a phase transition, but the photon stiffness is zero everywhere (the exhibited data reflects statistical noise in the measurement), and so both phases are confined. As we argue in the text, at large $\lambda$ we have the fractionalized Faraday lines phase, while the FFL-CFL transition is Ising-like.}
\label{cutC}
\end{figure}

The square symbols in Fig.~\ref{phase} mark points where we observed peaks in the specific heat, indicating a phase transition. We also obtained more detailed data on the dashed lines marked A, B, and C, and we present this data in Figs.~\ref{cutA}, \ref{cutB}, and \ref{cutC} respectively. 

Figure \ref{cutA} shows a transition between the Conventional Faraday Lines (CFL) and Coulomb phases. In the top panel we see that the peak in the specific heat grows rapidly as a function of system size, indicating a first-order phase transition. In the bottom panel we show the photon stiffness, which as expected is non-zero only in the Coulomb phase. The Coulomb to CFL transition can be thought of as a condensation of monopoles. The monopole fields can be described by a Higgs theory in (3+1)D, and such a theory is indeed expected to have a first-order transition.\cite{HLM,*ColemanWeinberg}

Figure \ref{cutB} shows a transition between the Fractionalized Faraday Lines (FFL) and Coulomb phases. As in the case above, our data shows that the transition is first-order and that the phase at large $\kappa$ is the Coulomb phase while the phase at small $\kappa$ is confined. The Coulomb-FFL transition can be thought of as a condensation of pairs of monopoles, and can be described by a Higgs theory of ``paired-monopole'' fields.  It is therefore not surprising that the transition also has the properties of a Higgs theory, and also that it takes place at a value of $\kappa$ approximately one-quarter that of the Coulomb-CFL transition at $\lambda=0$.

Finally, Fig.~\ref{cutC} shows a transition between the CFL and FFL phases. We can see from the specific heat data that there is indeed a phase transition here; however, from the photon stiffness we can see that both phases are confined. The specific heat peaks grow very slowly with the system size, indicating a second-order transition. In the next section we will argue that the transition is Ising-like, with mean-field critical indices consistent with our measurements.

\section{Explicit Demonstration of Fractionalization of Faraday lines and Properties of the Phase}

We now use an exact change of variables to rewrite the partition sum in Eq.~(\ref{action1}) in terms of gapped excitations of the FFL phase, which will make the properties of the phase clear. This procedure is valid for all space-time dimensions. We start by writing:
\begin{equation}
\uu_{\mu\nu}(r) = N m_{\mu\nu}(r) + \ell_{\mu\nu}(r) ~,
\end{equation}
where $m_{\mu\nu}(r)$ runs over arbitrary integer values while $\ell_{\mu\nu}(r)$ runs over integers $0, 1, \dots, N-1$.  The $\lambda$ term does not depend on $m_{\mu\nu}(r)$, and we can perform summation over these variables and obtain:
\begin{eqnarray}
&&Z = \int_0^{2\pi} \!\!\! Da_\mu(r) \!\!\! \sum_{\ell_{\mu\nu}(r) = 0}^{N-1}\!\!\!
\exp\Bigg( \lambda \!\!\! \sum_{r,~\sigma < \mu < \nu} \!\!\! \cos\left[\frac{2\pi \tilde{\cQ}_{\sigma\mu\nu}(r)}{N} \right]\\
&&- \!\!\! \sum_{r,~ \mu < \nu} \!\! V_{\rm Villain}\left[\left(\nabla_\mu \frac{a_\nu}{N} - \nabla_\nu \frac{a_\mu}{N}\right)(r) - \frac{2\pi \ell_{\mu\nu}(r)}{N}; \kappa N^2 \right] \Bigg).\nonumber
\label{Ztildea1}
\end{eqnarray}
Here $\tilde{\cQ}_{\sigma\mu\nu}(r)$ is given by Eq.~(\ref{Qsmn}) with $\uu_{\mu\nu}(r)$ replaced by $\ell_{\mu\nu}(r)$, and all arithmetic with the latter is understood to be modulo $N$ (also everywhere below).  

We now wish to replace the $a_\mu(r)$ variables with variables called $\tilde{a}_\mu(r)$, which satisfy the following condition:
\begin{equation}
(e^{i \tilde{a}_\mu(r)})^N = e^{ia_\mu(r)}.
\label{aa}
\end{equation}
Recalling that $e^{ia_\mu(r)}$ is the creation operator for a segment of a Faraday line, we see that $e^{i\atil}$ creates a segment of a fractionalized Faraday line carrying electric field strength of $1/N$ of the microscopic unit. The replacement $\tilde{a}_\mu(r) \equiv a_\mu(r)/N$ satisfies Eq.~(\ref{aa}) but leaves us with an $\atil$ that has a different integration range than $\ab$. Resolving this problem in a naive way by simply extending the range of integration leads to gauge-like redundancies which obscure some of the physics we are interested in.

To find variables $\atil$ which satisfy Eq.~(\ref{aa}) and avoid such redundancy problems, we take the following approach.  Working on a lattice with periodic boundary conditions, we can divide all configurations of $\ell_{\mu\nu}(r)$ into classes characterized by the fluxes $\tilde{\cQ}_{\sigma\mu\nu}(r)$ out of all elementary cubes, as well as fluxes through some fixed two-dimensional surfaces wrapping around the system with the periodic connectedness.  For example, we can take such surfaces which pass through the origin $(0, 0, \dots, 0)$ and define for each pair of directions $\hat{\mu}$ and $\hat{\nu}$ spanned by coordinates $0 \leq x_\mu \leq L_\mu - 1$ and $0 \leq x_\nu \leq L_\nu - 1$:
\begin{equation}
W_{\mu\nu} \equiv \sum_{x_\mu = 0}^{L_\mu - 1} \sum_{x_\nu = 0}^{L_\nu - 1} \ell_{\mu\nu}(0, \dots, 0, x_\mu, 0, \dots, 0, x_\nu, 0, \dots, 0) ~. \label{Wmn}
\end{equation}
We can check that
\begin{eqnarray}
\sum_{\ell_{\mu\nu}(r)} [\dots ] = \!\!\! && {\sum_{\tilde{\cQ}_{\sigma\mu\nu}(r), W_{\mu\nu}}}^\prime \sum_{v_\mu(r)} \\
&& [\ell_{\mu\nu}(r) = \ell^{(0)}_{\mu\nu}(r) + (\nabla_\mu v_\nu - \nabla_\nu v_\mu)(r)], ~~~~~~~
\label{sumell}
\end{eqnarray}
where $\ell^{(0)}_{\mu\nu}(r)$ is a fixed member of the class described by $\tilde{\cQ}_{\sigma\mu\nu}(r)$ and $W_{\mu\nu}$.  All summations are over integers $0, 1, \dots, N-1$ and all equations are modulo $N$ (i.e., the fields are elements of the additive group $\mathbb{Z}_N$).  To be precise, we can argue that from all links of the hyper-cubic lattice, we can select a subset of links such that we can take $v_\mu(r)$ as independent variables on these links, while $v_\mu(r) = 0$ on all the other links.  We can also argue that the original CQED theory can be equivalently formulated using compact gauge fields $a_\mu(r)$ that are non-zero only on exactly the same links as the independent $v_\mu(r)$.  We will assume this implicitly in all manipulations below.

The primed sum over $\tilde{\cQ}_{\sigma\mu\nu}(r)$ and $W_{\mu\nu}$ in Eq.~(\ref{sumell}) is over all such modulo-$N$ integers that can be derived from some $\ell_{\mu\nu}(r)$ via Eqs.~(\ref{Qsmn}) and (\ref{Wmn}).  We can also argue that such allowed $\tilde{\cQ}_{\sigma\mu\nu}(r)$ and $W_{\mu\nu}$ are equivalently described as independent $\tilde{\cQ}_{\sigma\mu\nu}(r)$ and $W_{\mu\nu}$ but with constraints on $\tilde{\cQ}_{\sigma\mu\nu}(r)$:
\begin{eqnarray}
&& \nabla_\rho \tilde{\cQ}_{\sigma\mu\nu}
- \nabla_\sigma \tilde{\cQ}_{\mu\nu\rho}
+ \nabla_\mu \tilde{\cQ}_{\nu\rho\sigma}
- \nabla_\nu \tilde{\cQ}_{\rho\sigma\mu} = 0 ~, \label{Qconstraints}\\
&& \sum_{x_\sigma = 0}^{L_\sigma - 1}
\sum_{x_\mu = 0}^{L_\mu - 1}
\sum_{x_\nu = 0}^{L_\nu - 1}
\tilde{\cQ}_{\sigma\mu\nu}(\dots, x_\sigma, \dots, x_\mu, \dots, x_\nu, \dots) = 0 ~, \nonumber
\end{eqnarray}
where in the last line all coordinates other than $x_\sigma$, $x_\mu$, and $x_\nu$ are fixed.  Note that while such $\tilde{\cQ}_{\sigma\mu\nu}(r)$ and $W_{\mu\nu}$ can be viewed as independent, the $\ell^{(0)}_{\mu\nu}(r)$ in Eq.~(\ref{sumell}) will depend on both; for each allowed $\tilde{\cQ}_{\sigma\mu\nu}(r)$ and $W_{\mu\nu}$ we choose some $\ell^{(0)}_{\mu\nu}(r)$ and treat it as a fixed function of $\tilde{\cQ}_{\sigma\mu\nu}(r)$ and $W_{\mu\nu}$.

Inserting Eq.~(\ref{sumell}) into the partition sum Eq.~(\ref{Ztildea1}), we note that $a_\mu(r)$ and $v_\mu(r)$ appear in a combination $a_\mu(r) - 2\pi v_\mu(r)$; hence we define
\begin{equation}
\tilde{a}_\mu(r) \equiv \frac{a_\mu(r)}{N} - \frac{2\pi v_\mu(r)}{N},
\label{tildea2}
\end{equation}
 which satisfies Eq.~(\ref{aa}).  Furthermore, integrating $a_\mu(r)$ over $[0, 2\pi)$ and summing $v_\mu(r)$ over $0, 1, \dots, N-1$ is equivalent to integrating $\tilde{a}_\mu(r)$ over $[0, 2\pi)$.  The partition sum becomes
\begin{widetext}
\begin{eqnarray}
Z \!=\! \int_0^{2\pi} \!\!\!\!\!\! D\tilde{a}_\mu(r) {\sum_{\tilde{\cQ}_{\sigma\mu\nu}(r)}}^\prime \sum_{W_{\mu\nu}}
\exp\left( \lambda \!\!\!\!\!\sum_{r,~\sigma < \mu < \nu}\!\!\! \cos\left[\frac{2\pi \tilde{\cQ}_{\sigma\mu\nu}(r)}{N} \right] 
- \!\!\! \sum_{r,~ \mu < \nu} \!\!\! V_{\rm Villain}\left[(\nabla_\mu \tilde{a}_\nu - \nabla_\nu \tilde{a}_\mu)(r) - \frac{2\pi \ell^{(0)}_{\mu\nu}(r)}{N}; \kappa N^2 \right] \right),
\label{Ztildea2}
\end{eqnarray}
\end{widetext}
where the primed sum over $\tilde{\cQ}_{\sigma\mu\nu}$ signifies the above-mentioned constraints on these variables.

At this point, the theory has a compact gauge field $\tilde{a}_\mu(r) \in [0, 2\pi)$ coupled to a $Z_N$ rank-2 field $\ell_{\mu\nu}(r)$.  In the absence of the $\tilde{a}_\mu(r)$ field, such a rank-2 theory undergoes a ``rank-2 confinement/deconfinement'' transition at some coupling strength $\lambda_c$ (assuming space-time dimensionality greater or equal to four). 
In the (3+1)D case, this transition has Ising-like critical exponents, which implies that the heat capacity peak should increase very slowly with the system size,\cite{isingnote,IsingExp} and our results in Fig.~\ref{cutC} are in agreement with this. In general dimension, adding a gapped rank-1 field $\tilde{a}_\mu(r)$---i.e., introducing only a small ``line dynamics'' parameter $\kappa$---will not affect the rank-2 confinement/deconfinement transition;\cite{LipsteinGerbes,JohnstonGerbes} however, it will destroy the higher-rank generalization of the Wilson loop diagnostics,\cite{LipsteinGerbes} similarly to what happens when adding a dynamical matter field to a lattice gauge theory.\cite{FradkinShenker79}  In the rank-2 deconfined phase at large $\lambda$ and small $\kappa$, the $\tilde{a}_\mu(r)$ fields represent true gapped line excitations---fractionalized Faraday lines carrying $1/N$ of the elementary electric field strength.

To bring out the fractionalized Faraday lines explicitly, we can Poisson-resum the Villain potential as in Eq.~(\ref{Villain}), introducing integer-valued placket variables $\tilde{F}_{\mu\nu}(r)$ for each placket. We can then perform integration over the $\tilde{a}_\mu(r)$ variables to get:
\begin{widetext}
\begin{eqnarray}
Z = {\sum_{\tilde{F}_{\mu\nu}(r)}}^\prime
{\sum_{\tilde{\cQ}_{\sigma\mu\nu}(r)}}^\prime \sum_{W_{\mu\nu}} 
\exp\left( \lambda \!\!\!\sum_{r,~\sigma < \mu < \nu} \!\!\!\cos\left[\frac{2\pi \tilde{\cQ}_{\sigma\mu\nu}(r)}{N} \right] 
- \sum_{r,~ \mu < \nu} \frac{\tilde{F}_{\mu\nu}(r)^2}{2 \kappa N^2}
- i \frac{2\pi}{N} \sum_{r,~ \mu < \nu} \tilde{F}_{\mu\nu}(r) \ell^{(0)}_{\mu\nu}(r) \right) ~,
\label{allgapped}
\end{eqnarray}
\end{widetext}
where the primed sum over $\tilde{F}_{\mu\nu}(r)$ denotes constraints
\begin{equation}
\sum_\nu \nabla_\nu \tilde{F}_{\mu\nu} = 0 ~.
\label{Fconstraints}
\end{equation}
We can view this final form as a representation in terms of gapped excitations of the phase with fractionalized Faraday lines realized at small $\kappa$ and large $\lambda$, assuming that the total space-time dimension is greater than or equal to four so that this phase exists.  The integer-valued $\tilde{F}_{\mu\nu}(r)$ variables satisfying Eq.~(\ref{Fconstraints}) represent worldsheets (i.e., space-time ``history'') of the fractionalized quantum Faraday lines.  On the other hand, the $Z_N$-valued rank-3 objects $\tilde{Q}_{\sigma\mu\nu}(r)$ satisfying Eq.~(\ref{Qconstraints}) in general dimensionality represent the space-time history of quantum surfaces, while in (3+1) dimensions they are equivalent to conserved $Z_N$-valued currents representing worldines of quantum particles, see Eq.~(\ref{Jdef}), and in (4+1) dimensions they are equivalent to $Z_N$-valued quantum lines.  There is a large penalty for having either of these objects, so they are gapped in this phase.

The last term in the action encodes statistical interaction between the fractionalized quantum Faraday lines and the $Z_N$ quantum surfaces.
As an example, in (3+1)D where the $Z_N$ objects can be equivalently viewed as representing quantum particles, when an elementary $Z_N$ particle goes around an elementary fractionalized Faraday line, there is a Berry phase of $2\pi/N$.  The $Z_N$ particle can be viewed as a single monopole in the presence of the $N$-tupled monopole condensate, and the statistical interaction with fractionalized Faraday lines can be readily anticipated from our discussion in the Introduction. 

Besides the gapped excitations, we also see the appearance of topological sectors described by $W_{\mu\nu} = 0, 1, \dots, N-1$ for each pair of periodic directions $\hat{\mu}$ and $\hat{\nu}$. We remark that in the present approach we did not add any redundant degrees of freedom and all ``counting'' is precise, so the sectors are real.  We will now argue that in the regime of small $\kappa$ and large $\lambda$, these sectors correspond to a topological ground state degeneracy in such a phase with fractionalized Faraday lines.

First let us focus on the $Z_N$ rank-2 system ignoring the fractionalized Faraday lines.  Note that the $\lambda$ term in the action does not depend on $W_{\mu\nu}$.  While we can change the value of $W_{\mu\nu}$ by changing a single $\ell_{\mu\nu}(r)$ on the defining surface in Eq.~(\ref{Wmn}), this also changes the nearby $\tilde{\cQ}_{\sigma\mu\nu}(r)$.  To change the value of $W_{\mu\nu}$ without changing any $\tilde{\cQ}_{\sigma\mu\nu}(r)$, we need to change by the same amount all $\ell_{\mu\nu}(r)$ at some fixed $x_\mu$ and $x_\nu$, i.e., we need $\prod_{\sigma \neq \mu, \nu} L_\sigma$ such local placket variable changes.  Our interpretation of this is that on a $d$-dimensional spatial torus of volume $L^d$ and in the absence of the fractionalized Faraday lines, the tunneling (mixing) between the above sectors will be exponentially small in $L^{d-2}$.

Let us now include the fractionalized Faraday lines.  They see the different sectors only through the global ``fluxes'' of  $\ell^{(0)}_{\mu\nu}(r)$, i.e., only when the fractionalized Faraday lines sweep in their motion the entire $L_\mu \times L_\nu$ wrapping surface.  Hence when the fractionalized Faraday lines are gapped, the splitting (diagonal energy difference) between the different $W_{\mu\nu}$ sectors will be exponentially small in $L_\mu \times L_\nu$.  

Considering all effects together, on the $d$-dimensional spatial torus we expect $N^{d(d-1)/2}$ nearly degenerate ground states with splittings which are exponentially small in the square of the linear size of the system. 

We found these results by considering a classical action in $(d+1)$ space-time dimensions. The same topological degeneracy was found in Ref.~\onlinecite{Yoshida} for a quantum lattice Hamiltonian in $d$ spatial dimensions which is an extension of the toric code where, as in our model, the degrees of freedom live on the plackets of a (hyper)cubic lattice.  This Hamiltonian has ``star'' terms associated with links and ``placket'' terms associated with three-dimensional cubes, as in our model, and is labeled ``$(d, 2)$'' in Ref.~\onlinecite{Yoshida}, Appendix B.2.  The Euclidean path integral for this Hamiltonian in the sector with fixed star terms gives the same space-time action as our $Z_N$ rank-2 system, just like the familiar Kitaev's toric code in the sector with fixed star terms associated with sites gives the classical Ising gauge theory.  Since it is the deconfined rank-2 system that is responsible for the topological degeneracy, the connection with the Hamiltonian in Ref.~\onlinecite{Yoshida} confirms our analysis of the topological degeneracy, even though the broader setting here realizing fractionalized Faraday lines in a generalized CQED model is different from the setting in Ref.~\onlinecite{Yoshida}, which is considering topological phases of short-range interacting spins.

\section{Discussion}
Topological field theories have been an active area of research for a long time, and recently progress has been made by generalizing ideas from topological phases of bosons to propose topological phases of gauge theories.\cite{GukovKapustin, KapustinThorngren, McGreevy} In this paper we present a lattice CQED model realizing condensation of bound states of multiple monopoles and producing a phase with emergent intrinsic topological order, analogous to the way condensing multiple vortices in a boson system can give a fractionalized Mott insulator. The phase we find contains gapped excitations which are fractionalized Faraday lines and additional excitations which are quantum surfaces in spatial dimensions above four, but can be viewed as quantum lines or quantum particles in four or three spatial dimensions respectively. 
These excitations have statistical interactions with the fractionalized Faraday lines encoded by the last term in the final representation Eq.~(\ref{allgapped}), or equivalently by the minimal coupling of the 1-form $\tilde{a}_\mu(r)$ to the 2-form $ \ell_{\mu\nu}(r)$ in the representation Eq.~(\ref{Ztildea2}).
Thus, our model is also an example of a lattice Gerbe theory\cite{LipsteinGerbes, JohnstonGerbes} emerging as an effective field theory description of the CQED phase with fractionalized Faraday lines, similar to how lattice gauge theories can emerge as descriptions of fractionalized phases of bosons with short-ranged interactions.  

Another set of examples of novel phases of lattice gauge systems that are higher-rank analogs of symmetry-protected topological (non-fractionalized) phases of bosons and symmetry-enriched topological (fractionalized) phases of bosons can be found in the Appendix of Ref.~\onlinecite{SO34D}, where we construct models with CQED$\times$boson symmetries realizing condensates of bound states of monopoles and bosons.  
More broadly, we think that the idea of condensing bound states of topological defects and symmetry-charged objects\cite{Geraedts2013} can yield precise models of emergent topological phases in many other lattice gauge theory systems.\cite{GukovKapustin, KapustinThorngren, McGreevy, KeyserlingkBurnell2014}

\acknowledgments
We would like to thank M.~Metlitski, M.~P.~A.~Fisher, F.~Burnell, A.~Kapustin, C.~von~Keyserlingk, J.~Preskill, T.~Senthil, and A.~Vishwanath for many inspiring discussions.  This research is supported by the National Science Foundation through grant DMR-1206096, and by the Caltech Institute of Quantum Information and Matter, an NSF Physics Frontiers Center with support of the Gordon and Betty Moore Foundation.

\bibliography{bib4fractfaraday}

\begin{thebibliography}{10}%
\makeatletter
\providecommand \@ifxundefined [1]{%
 \ifx #1\undefined \expandafter \@firstoftwo
 \else \expandafter \@secondoftwo
\fi
}%
\providecommand \@ifnum [1]{%
 \ifnum #1\expandafter \@firstoftwo
 \else \expandafter \@secondoftwo
\fi
}%
\providecommand \enquote [1]{``#1''}%
\providecommand \bibnamefont  [1]{#1}%
\providecommand \bibfnamefont [1]{#1}%
\providecommand \citenamefont [1]{#1}%
\providecommand\href[0]{\@sanitize\@href}%
\providecommand\@href[1]{\endgroup\@@startlink{#1}\endgroup\@@href}%
\providecommand\@@href[1]{#1\@@endlink}%
\providecommand \@sanitize [0]{\begingroup\catcode`\&12\catcode`\#12\relax}%
\@ifxundefined \pdfoutput {\@firstoftwo}{%
 \@ifnum{\z@=\pdfoutput}{\@firstoftwo}{\@secondoftwo}%
}{%
 \providecommand\@@startlink[1]{\leavevmode\special{html:<a href="#1">}}%
 \providecommand\@@endlink[0]{\special{html:</a>}}%
}{%
 \providecommand\@@startlink[1]{%
  \leavevmode
  \pdfstartlink
   attr{/Border[0 0 1 ]/H/I/C[0 1 1]}%
   user{/Subtype/Link/A<</Type/Action/S/URI/URI(#1)>>}%
  \relax
 }%
 \providecommand\@@endlink[0]{\pdfendlink}%
}%
\providecommand \url  [0]{\begingroup\@sanitize \@url }%
\providecommand \@url [1]{\endgroup\@href {#1}{\urlprefix}}%
\providecommand \urlprefix [0]{URL }%
\providecommand \Eprint[0]{\href }%
\@ifxundefined \urlstyle {%
  \providecommand \doi [1]{doi:\discretionary{}{}{}#1}%
}{%
  \providecommand \doi [0]{doi:\discretionary{}{}{}\begingroup
  \urlstyle{rm}\Url }%
}%
\providecommand \doibase [0]{http://dx.doi.org/}%
\providecommand \Doi[1]{\href{\doibase#1}}%
\providecommand \bibAnnote [3]{%
  \BibitemShut{#1}%
  \begin{quotation}\noindent
    \textsc{Key:}\ #2\\\textsc{Annotation:}\ #3%
  \end{quotation}%
}%
\providecommand \bibAnnoteFile [2]{%
  \IfFileExists{#2}{\bibAnnote {#1} {#2} {\input{#2}}}{}%
}%
\providecommand \typeout [0]{\immediate \write \m@ne }%
\providecommand \selectlanguage [0]{\@gobble}%
\providecommand \bibinfo [0]{\@secondoftwo}%
\providecommand \bibfield [0]{\@secondoftwo}%
\providecommand \translation [1]{[#1]}%
\providecommand \BibitemOpen[0]{}%
\providecommand \bibitemStop [0]{}%
\providecommand \bibitemNoStop [0]{.\EOS\space}%
\providecommand \EOS [0]{\spacefactor3000\relax}%
\providecommand \BibitemShut [1]{\csname bibitem#1\endcsname}%
\bibitem{KapustinThorngren}%
  \BibitemOpen
  \bibfield{author}{%
  \bibinfo {author} {\bibfnamefont{A.}~\bibnamefont{Kapustin}}\ and\ \bibinfo
  {author} {\bibfnamefont{R.}~\bibnamefont{Thorngren}},\ }%
  \bibfield{journal}{%
  \bibinfo {journal} {arXiV/hep-th},\ \bibinfo {pages} {1309.4721}}%
   (\bibinfo {year} {2013})%
  \bibAnnoteFile{NoStop}{KapustinThorngren}%
\bibitem{GukovKapustin}%
  \BibitemOpen
  \bibfield{author}{%
  \bibinfo {author} {\bibfnamefont{S.}~\bibnamefont{Gukov}}\ and\ \bibinfo
  {author} {\bibfnamefont{A.}~\bibnamefont{Kapustin}},\ }%
  \bibfield{journal}{%
  \bibinfo {journal} {arXiv/hep-th},\ \bibinfo {pages} {1307.4793}}%
   (\bibinfo {year} {2013})%
  \bibAnnoteFile{NoStop}{GukovKapustin}%
\bibitem{BalentsFisherNayak99}%
  \BibitemOpen
  \bibfield{author}{%
  \bibinfo {author} {\bibfnamefont{L.}~\bibnamefont{Balents}}, \bibinfo
  {author} {\bibfnamefont{M.~P.~A.}\ \bibnamefont{Fisher}},\ and\ \bibinfo
  {author} {\bibfnamefont{C.}~\bibnamefont{Nayak}},\ }%
  \bibfield{journal}{%
  \bibinfo {journal} {Phys. Rev. B}\ }%
  \textbf{\bibinfo {volume} {60}},\ \bibinfo {pages} {1654} (\bibinfo {year}
  {1999})%
  \bibAnnoteFile{NoStop}{BalentsFisherNayak99}%
\bibitem{SenthilFisher00}%
  \BibitemOpen
  \bibfield{author}{%
  \bibinfo {author} {\bibfnamefont{T.}~\bibnamefont{Senthil}}\ and\ \bibinfo
  {author} {\bibfnamefont{M.~P.~A.}\ \bibnamefont{Fisher}},\ }%
  \bibfield{journal}{%
  \bibinfo {journal} {Phys. Rev. B}\ }%
  \textbf{\bibinfo {volume} {62}},\ \bibinfo {pages} {7850} (\bibinfo {year}
  {2000})%
  \bibAnnoteFile{NoStop}{SenthilFisher00}%
\bibitem{Polyakov}%
  \BibitemOpen
  \bibfield{author}{%
  \bibinfo {author} {\bibfnamefont{A.~M.}\ \bibnamefont{Polyakov}},\ }%
  \emph{\bibinfo {title} {Gauge Fields and Strings}}\ (\bibinfo {publisher}
  {Hardwood Academic Publishers},\ \bibinfo {year} {1987})%
  \bibAnnoteFile{NoStop}{Polyakov}%
\bibitem{footnote1}%
  \BibitemOpen
  \bibinfo {journal} {In the Monte Carlo we need to condense objects of $N$
  monopoles, and the amount of Monte Carlo time required to do this increases
  with $N$. We have found evidence for the fractionalized phase in Monte Carlo
  up to $N=5$.}%
  \bibAnnoteFile{Stop}{footnote1}%
\bibitem{HLM}%
  \BibitemOpen
\bibfield{journal}{%
    }%
  \bibfield{author}{%
  \bibinfo {author} {\bibfnamefont{B.~I.}\ \bibnamefont{Halperin}}, \bibinfo
  {author} {\bibfnamefont{T.~C.}\ \bibnamefont{Lubensky}},\ and\ \bibinfo
  {author} {\bibfnamefont{S.-k.}\ \bibnamefont{Ma}},\ }%
  \bibfield{journal}{%
  \bibinfo {journal} {Phys. Rev. Lett.}\ }%
  \textbf{\bibinfo {volume} {32}},\ \bibinfo {pages} {292} (\bibinfo {year}
  {1974})%
  \bibAnnoteFile{NoStop}{HLM}%
\bibitem{ColemanWeinberg}%
  \BibitemOpen
  \bibfield{author}{%
  \bibinfo {author} {\bibfnamefont{S.}~\bibnamefont{Coleman}}\ and\ \bibinfo
  {author} {\bibfnamefont{E.}~\bibnamefont{Weinberg}},\ }%
  \bibfield{journal}{%
  \bibinfo {journal} {Phys. Rev. D}\ }%
  \textbf{\bibinfo {volume} {7}},\ \bibinfo {pages} {1888} (\bibinfo {year}
  {1973})%
  \bibAnnoteFile{NoStop}{ColemanWeinberg}%
\bibitem{isingnote}%
  \BibitemOpen
  \emph{\bibinfo {title} {{\rm Mean field predicts a step discontinuity for the
  specific heat, while scaling theory predicts $\sqrt[3]{\log L}$
  divergence.\cite{IsingExp} Our data is consistent with both predictions (we
  are not able to access large enough system sizes to distinguish between
  them).}}}%
  \bibAnnoteFile{Stop}{isingnote}%
\bibitem{IsingExp}%
  \BibitemOpen
  \bibfield{author}{%
  \bibinfo {author} {\bibfnamefont{P.~H.}\ \bibnamefont{Lundow}}\ and\ \bibinfo
  {author} {\bibfnamefont{K.}~\bibnamefont{Markstr\"om}},\ }%
  \bibfield{journal}{%
  \bibinfo {journal} {Phys. Rev. E}\ }%
  \textbf{\bibinfo {volume} {80}},\ \bibinfo {pages} {031104} (\bibinfo {year}
  {2009})%
  \bibAnnoteFile{NoStop}{IsingExp}%
\bibitem{LipsteinGerbes}%
  \BibitemOpen
  \bibfield{author}{%
  \bibinfo {author} {\bibfnamefont{A.~E.}\ \bibnamefont{Lipstein}}\ and\
  \bibinfo {author} {\bibfnamefont{R.~A.}\ \bibnamefont{Reid-Edwards}},\ }%
  \bibfield{journal}{%
  \bibinfo {journal} {arXiv/hep-th},\ \bibinfo {pages} {1404.2634}}%
   (\bibinfo {year} {2014})%
  \bibAnnoteFile{NoStop}{LipsteinGerbes}%
\bibitem{JohnstonGerbes}%
  \BibitemOpen
  \bibfield{author}{%
  \bibinfo {author} {\bibfnamefont{D.~A.}\ \bibnamefont{Johnston}},\ }%
  \bibfield{journal}{%
  \bibinfo {journal} {arXiv/hep-th},\ \bibinfo {pages} {1405.7890}}%
   (\bibinfo {year} {2014})%
  \bibAnnoteFile{NoStop}{JohnstonGerbes}%
\bibitem{FradkinShenker79}%
  \BibitemOpen
  \bibfield{author}{%
  \bibinfo {author} {\bibfnamefont{E.}~\bibnamefont{Fradkin}}\ and\ \bibinfo
  {author} {\bibfnamefont{S.~H.}\ \bibnamefont{Shenker}},\ }%
  \bibfield{journal}{%
  \bibinfo {journal} {Phys. Rev. D}\ }%
  \textbf{\bibinfo {volume} {19}},\ \bibinfo {pages} {3682} (\bibinfo {year}
  {1979})%
  \bibAnnoteFile{NoStop}{FradkinShenker79}%
\bibitem{Yoshida}%
  \BibitemOpen
  \bibfield{author}{%
  \bibinfo {author} {\bibfnamefont{B.}~\bibnamefont{Yoshida}},\ }%
  \bibfield{journal}{%
  \bibinfo {journal} {Annals of Physics}\ }%
  \textbf{\bibinfo {volume} {326}},\ \bibinfo {pages} {2566 } (\bibinfo {year}
  {2011})%
  \bibAnnoteFile{NoStop}{Yoshida}%
\bibitem{McGreevy}%
  \BibitemOpen
  \bibfield{author}{%
  \bibinfo {author} {\bibfnamefont{S.~M.}\ \bibnamefont{Kravec}}\ and\ \bibinfo
  {author} {\bibfnamefont{J.}~\bibnamefont{McGreevy}},\ }%
  \bibfield{journal}{%
  \bibinfo {journal} {Phys. Rev. Lett.}\ }%
  \textbf{\bibinfo {volume} {111}},\ \bibinfo {pages} {161603} (\bibinfo {year}
  {2013})%
  \bibAnnoteFile{NoStop}{McGreevy}%
\bibitem{SO34D}%
  \BibitemOpen
  \bibfield{author}{%
  \bibinfo {author} {\bibfnamefont{S.~D.}\ \bibnamefont{Geraedts}}\ and\
  \bibinfo {author} {\bibfnamefont{O.~I.}\ \bibnamefont{Motrunich}},\ }%
  \bibfield{journal}{%
  \bibinfo {journal} {arXiv:cond-mat},\ \bibinfo {pages} {1408.1096}}%
   (\bibinfo {year} {2014})%
  \bibAnnoteFile{NoStop}{SO34D}%
\bibitem{Geraedts2013}%
  \BibitemOpen
  \bibfield{author}{%
  \bibinfo {author} {\bibfnamefont{S.~D.}\ \bibnamefont{Geraedts}}\ and\
  \bibinfo {author} {\bibfnamefont{O.~I.}\ \bibnamefont{Motrunich}},\ }%
  \bibfield{journal}{%
  \bibinfo {journal} {Annals of Physics}\ }%
  \textbf{\bibinfo {volume} {334}},\ \bibinfo {pages} {288 } (\bibinfo {year}
  {2013})%
  \bibAnnoteFile{NoStop}{Geraedts2013}%
\bibitem{KeyserlingkBurnell2014}%
  \BibitemOpen
  \bibfield{author}{%
  \bibinfo {author} {\bibfnamefont{C.~W.}\ \bibnamefont{{von Keyserlingk}}}\
  and\ \bibinfo {author} {\bibfnamefont{F.~J.}\ \bibnamefont{{Burnell}}},\ }%
  \bibfield{journal}{%
  \bibinfo {journal} {ArXiv e-prints}}%
   (\bibinfo {year} {2014}),\
  \Eprint{http://arxiv.org/abs/1405.2988}{arXiv:1405.2988}%
  \bibAnnoteFile{NoStop}{KeyserlingkBurnell2014}%
\end{thebibliography}%

\end{document}